\documentclass{emulateapj}

\slugcomment{}

\shorttitle{The spectral signatures of LyC leakage}
\shortauthors{Zackrisson et al.}

\begin{document}

\title{The spectral evolution of the first galaxies. II.\\ Spectral signatures of Lyman continuum leakage from\\ galaxies in the reionization epoch}

\author{Erik Zackrisson\altaffilmark{1}$^*$, Akio K. Inoue\altaffilmark{2} \& Hannes Jensen\altaffilmark{1}}
\altaffiltext{*}{E-mail: ez@astro.su.se}
\altaffiltext{1}{The Oskar Klein Centre, Department of Astronomy, AlbaNova, Stockholm University, SE-106 91 Stockholm, Sweden}
\altaffiltext{2}{College of General Education, Osaka Sangyo University, 3-1-1, Nakagaito, Daito, Osaka 574-8530, Japan}

\begin{abstract}
The fraction of ionizing photons ($f_\mathrm{esc}$) that escape from $z\gtrsim 6$ galaxies is an important parameter when assessing the role of these objects in the reionization of the Universe, but the opacity of the intergalactic medium precludes a direct measurement of $f_\mathrm{esc}$ for individual galaxies at these epochs. We argue, that since $f_\mathrm{esc}$ regulates the impact of nebular emission on the spectra of galaxies, it should nonetheless be possible to indirectly probe $f_\mathrm{esc}$ well into the reionization epoch. As a first step, we demonstrate that by combining measurements of the rest-frame UV slope $\beta$ with the equivalent width of the $H\beta$ emission line, galaxies with very high Lyman continuum escape fractions ($f_\mathrm{esc}\geq 0.5$) should be identifiable up to $z\approx 9$ through spectroscopy with the upcoming James Webb Space Telescope (JWST). By targeting strongly lensed galaxies behind low-redshift galaxy clusters, JWST spectra of sufficiently good quality can be obtained for $M_{1500}\lesssim -16.0$ galaxies at $z\approx 7$ and for $M_{1500}\lesssim -17.5$ galaxies at $z\approx 9$. Dust-obscured star formation may complicate the analysis, but supporting observations with ALMA or the planned SPICA mission may provide useful constraints on the dust properties of these galaxies.  
\end{abstract}





\section{Introduction}
\label{intro}
The spectra of high-redshift quasars suggest that cosmic reionization was completed by $z\approx 6$ \citep{Fan et al., mortlock2011}, and recent measurements of the kinetic Sunyaev-Zeldovich effect constrain the duration of this process to $\Delta z\leq 7.9$ \citep{Zahn et al.}. The galaxy population at $z\gtrsim 6$ may in principle be sufficient to reionize the Universe \citep[e.g.][]{McLure et al.,Oesch et al.,Bouwens et al. b,Lorenzoni et al.,Grazian et al.,Finkelstein et al. b,Robertson et al. b}, but this hinges on the slope of the galaxy luminosity function at luminosities significantly below current detection thresholds, and on the fraction of hydrogen-ionizing photons that escape from galaxies into the intergalactic medium (IGM). The latter quantity, the Lyman-continuum (LyC) escape fraction $f_\mathrm{esc}$, can be directly measured at $z\lesssim 4$ \citep[e.g.][]{Steidel et al.,Shapley et al.,Bergvall et al. a,Iwata et al.,Bogosavljevic,Vanzella10,Siana et al.,Boutsia et al.,Nestor11,Vanzella12,Nestor13,Leitet et al.}. Observations of this type indicate an increase in the typical $f_\mathrm{esc}$ with redshift in the $z\approx 0$--3 interval \citep[][]{Inoue et al. b,Bergvall et al. b}, and data-constrained models for galaxy-dominated reionization also require a redshift evolution in $f_\mathrm{esc}$ \citep[e.g.][]{Kuhlen12,Fontanot et al.,Yue et al.,Ferrara & Loeb,Mitra et al.}. Simulations and theoretical arguments moreover suggest that $f_\mathrm{esc}$ may vary as a function of galaxy mass and star-formation activity \citep[e.g.][]{Razoumov & Sommer-Larsen,Yajima et al.,Fernandez & Shull,Conroy & Kratter}. 

At redshifts $z\gtrsim 4$, measurements of the rest frame LyC flux (at wavelengths $\leq 912$ \AA) are precluded by the opacity of the increasingly neutral IGM \citep{Inoue & Iwata}. To probe $f_\mathrm{esc}$ throughout the reionization epoch, indirect measurements are instead required. Some constraints on the {\it typical} $f_\mathrm{esc}$ in the reionization epoch may be obtained from fluctuations in the cosmic infrared background fluctuations \citep{Fernandez et al.}, or by combining the observed galaxy luminosity function with Ly$\alpha$ forest data \citep[e.g.][]{Finkelstein et al. b}, but it remains unclear whether and how $f_\mathrm{esc}$ may be assessed for individual galaxies at these redshifts. 
\begin{figure*}
\centering
\includegraphics[scale=0.7]{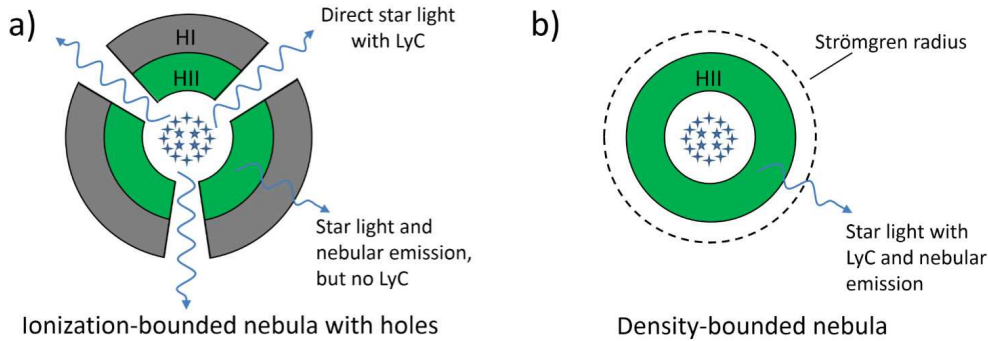}
\caption{Schematic illustrations of LyC escape mechanisms. A galaxy is here represented by a centrally concentrated ensemble of stars, surrounded by a photoionized HII region (green) and possibly an outer HI layer (gray). {\bf a)} A radiation-bounded HII region, in which holes in the ISM allow LyC photons to escape from the galaxy. {\bf b)} A density-bounded HII region, formed by a powerful starburst that photoionizes all the neutral gas in the galaxy without being able to form a complete Str\"omgren sphere (dashed line). \label{geometries}}
\end{figure*}

Here, we argue that since $f_\mathrm{esc}$ regulates the impact of nebular emission on the spectral energy distribution (SEDs) of galaxies, it should be possible to identify high-$f_\mathrm{esc}$ candidates from their rest-frame ultraviolet/optical SEDs at $z\gtrsim 6$, at least for objects with very high escape fractions ($f_\mathrm{esc}\gtrsim 0.5$). \citet{Ono et al.}, \citet{Bergvall et al. b} and \citet{Pirzkal et al. a,Pirzkal et al. b} have previously attempted to assess $f_\mathrm{esc}$ from  photometric data, but spectroscopic signatures are likely to produce more reliable results for individual targets. Recently, \citet{Jones et al.} presented a spectroscopic method to place upper limits on the LyC escape fraction of high-redshift galaxies using metal absorption lines, but unlike their method, the one proposed here should be able to place both upper and lower limits on the escape fraction. 

In Sect.~\ref{mechanisms}, we describe the two main mechanisms capable of producing LyC leakage in galaxies and introduce the geometries adopted in our subsequent modelling of these. In Sect.~\ref{signatures}, we present a simple spectral diagnostic that may be used to assess $f_\mathrm{esc}$ in the case of high-leakage objects, and discuss the role of metallicity, star formation history and dust attenuation on the relevant spectral features. Focusing on the capabilities of the 
{\it Near InfraRed Spectrograph (NIRSpec)} on the upcoming {\it James Webb Space Telescope (JWST)}, we derive the galaxy luminosity limits for the proposed method to estimate $f_\mathrm{esc}$ in Sect.~\ref{discussion}. A number of lingering problems with the proposed method are also discussed.  Sect.~\ref{summary} summarizes our findings. All our calculations are based on a $\Omega_\mathrm{M}=0.3$, $\Omega_\mathrm{\Lambda}=0.7$, $H_0=70$ km s$^{-1}$ Mpc$^{-1}$ cosmology, in rough agreement with the constraints set by the WMAP 9-year data combined with $H_0$ measurments and baryon acoustic oscillations \citep{Hinshaw et al.} and the first Planck results \citep{Ade et al.}. Whenever UV slopes and line equivalent widths are discussed, these quantities are given for rest-frame SEDs.

\section{Leakage mechanisms}
\label{mechanisms}
There are basically two different mechanisms that can cause LyC leakage from star-forming regions -- an radiation-bounded nebula with holes, and a density-bounded nebula (also known as a truncated Str\"omgren sphere). These two scenarios are schematically illustrated in Fig.~\ref{geometries}, in which our model galaxy is depicted as a centrally concentrated ensemble of stars surrounded by a single HII region. The first case (Fig.~\ref{geometries}a) corresponds to the situation when supernovae or stellar winds have opened up low-density channels in the neutral interstellar medium (ISM) through which LyC photons may escape without getting absorbed\footnote{``holes'' of this type are sometimes referred to as supernova chimneys or galactic fountains}. The second case (Fig.~\ref{geometries}b) corresponds to a situation when the LyC flux from a very powerful star-formation episode ``exhausts'' all the HI before a complete Str\"omgren sphere can form, thereby allowing LyC photons to escape into the IGM.

Galaxies are admittedly more complex than the simple, toy-model geometries depicted in Fig.~\ref{geometries} and contain a spatially extended ensemble of HII regions with different sizes and densities. Actual cases of LyC leakage are therefore likely to be due to mixtures of the two mechanisms. However, as we will demonstrate in Sect.~\ref{EW_beta}, the spectral diagnostics we propose are very similar for these two limiting cases, as long as dust effects can be ignored. 
Dust attenuation does, however, affect these two geometries differently, as discussed in Sect. \ref{attenuation}. The situation where LyC leakage is caused by high-velocity, LyC-emitting stars that venture far from the centres of galaxies \citep{Conroy & Kratter} is admittedly not well-represented by these simple geometries, but should effectively give rise to a situation similar to that in Fig.~\ref{geometries}b.

\section{Spectral diagnostics of Lyman continuum leakage}
\label{signatures}
\begin{figure*}
\plottwo{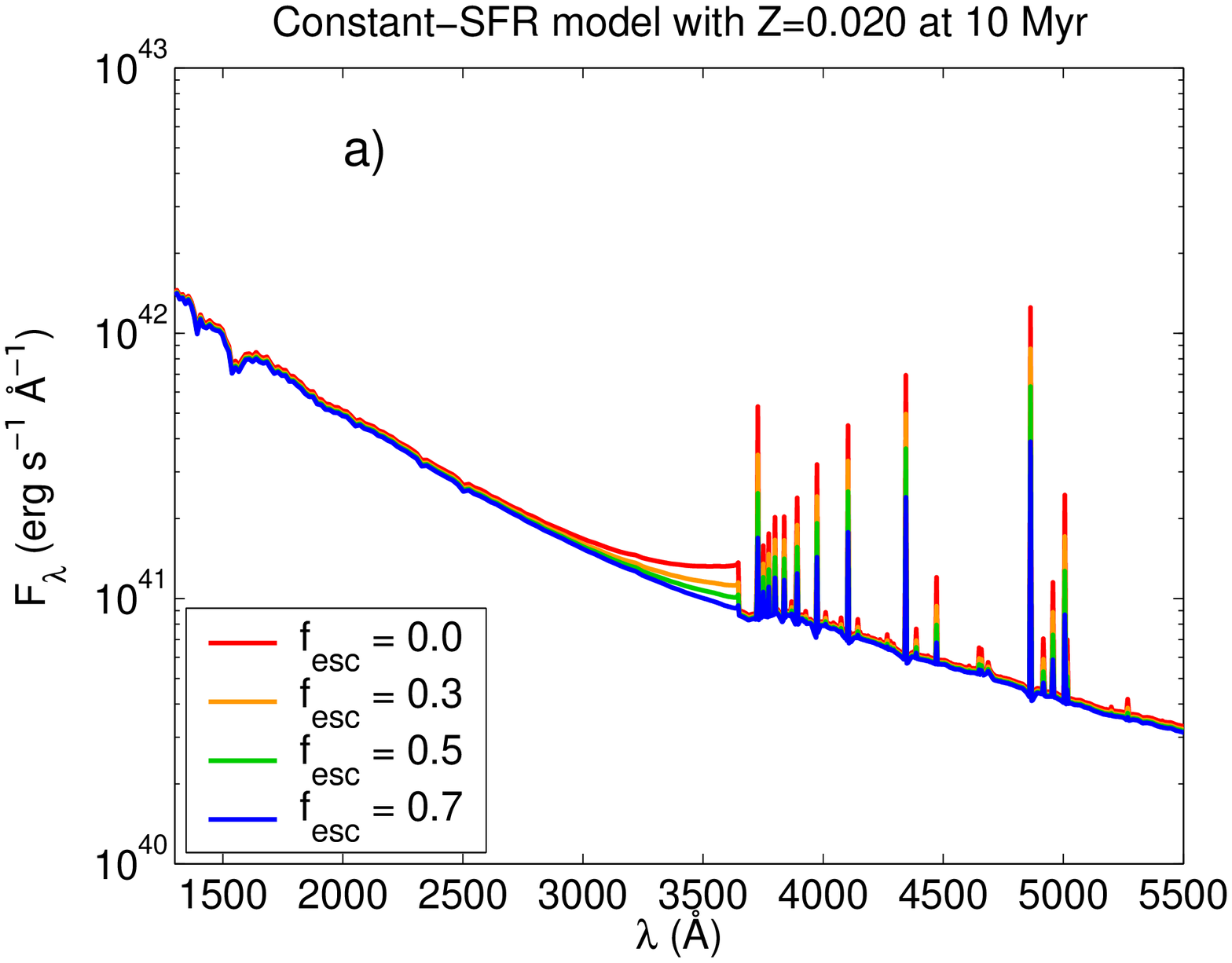}{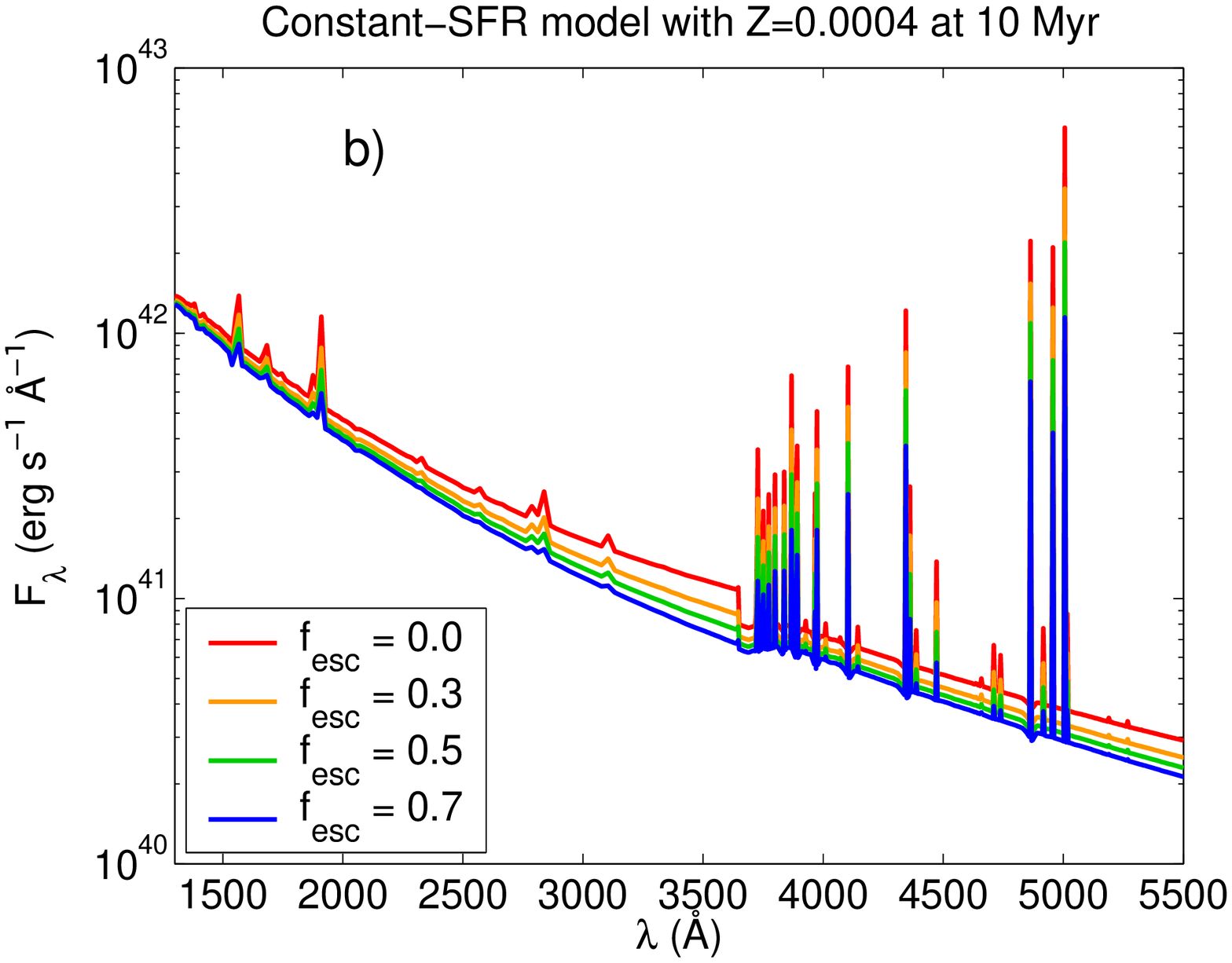}
\caption{The impact of the LyC escape fraction $f_\mathrm{esc}$ on the rest frame UV/optical SED of a young (age 10 Myr) starburst with stellar mass $10^9\ M_\odot$, a constant star formation rate and metallicity  $Z=0.020$ ({\bf a}) or $Z=0.0004$ ({\bf b}). The line colours represent different $f_\mathrm{esc}$: 0.0 (red), 0.3 (orange), 0.5 (green) and 0.7 (blue). As seen, both the strengths of the emission lines and the amplitude of the Balmer jump at 3646 \AA{} drop with increasing $f_\mathrm{esc}$. In the $Z=0.0004$ case, the entire slope of the UV continuum changes with $f_\mathrm{esc}$ (becomes bluer with increasing $f_\mathrm{esc}$), simply because the relative contribution of nebular emission to the overall SED is higher at this metallicity).
\label{spectra}}
\end{figure*}

\subsection{Model assumptions}
The SED model results presented in this paper are based on the {\it Yggdrasil} spectral synthesis code \citep{Zackrisson et al. b}, which mixes single-age stellar population spectra to simulate arbitrary star formation histories. These are used as input to the photoionization code Cloudy \citep{Ferland et al.}. The result is an age sequence of model SEDs which both stellar continuum emission, nebular continuum emission and nebular emission lines. In this paper, we adopt Starburst99 stellar SEDs generated using Padova-AGB stellar evolutionary tracks \citep{Vazquez & Leitherer} at metallicities $Z=0.0004$--0.020, assuming the \citet{Kroupa} universal stellar initial mass function (IMF) throughout the mass range 0.1--100 $M_\odot$. Since we are focusing on galaxies well into the reionization epoch, we limit the discussion to objects at $z\gtrsim 7$, with ages $\lesssim 7\times 10^8$ (roughly the age of the Universe at $z=7$). Population III galaxies (at zero or near-zero metallicities) may also exist at high redshifts \citep[e.g.][]{Stiavelli & Trenti} and are predicted to exhibit pronouned LyC leakage \citep{Johnson et al.}. While such objects likely have too low masses to contribute substantially to the reionization of the Universe, we briefly discuss models for such galaxies in Sect.~\ref{popIII}, using stellar SEDs from \citet{Raiter et al.}.

Yggdrasil assumes that the nebular component of the overall galaxy SED can be treated as originating from a single, spherical and isotropic HII region (as schematically depicted in Fig.~\ref{geometries}) with constant density and filling factor. In the case of radiation-bounded nebulae with holes (Fig.~\ref{geometries}a), model SEDs with different $f_\mathrm{esc}$ are generated by varying the covering factor of the nebula in Cloudy. In the case of density-bounded nebulae (Fig.~\ref{geometries}b), different $f_\mathrm{esc}$ are instead achieved by truncating the nebula at a fixed fraction of the theoretical Str\"omgren radius. In this paper, we adopt a constant hydrogen number density $n_{H}=100$ cm$^{-3}$, a filling factor of 0.01 and a gas metallicity identical to that of the stars. However, the primary diagnostics that we discuss (Balmer recombination lines and nebular continuum in the rest-frame UV/optical) are not sensitive to these parameter choices. We have moreover checked that our predictions for radiation-bounded nebulae with holes (Fig.~\ref{geometries}a) are in reasonable agreement with those of the \citet{Inoue c,Inoue d} model, which is based on a slightly different computational machinery.

\subsection{The impact of the LyC leakage of the SEDs of young galaxies}
Some of the ionizing photons produced by massive, hot stars are absorbed by the neutral hydrogen in the ISM and produce prominent HII regions. As a result, photons at rest-frame wavelengths $\lambda< 912$ \AA{} get reprocessed into emission lines and nebular continuum flux at longer wavelengths. A substantial fraction of the observed rest-frame UV/optical fluxes of young and/or star-forming galaxies is therefore expected to come from nebular emission rather than direct star light \citep[e.g.][]{{Zackrisson et al. a,Schaerer & de Barros a,Inoue d}}. Since the relative contribution from nebular emission to the overall galaxy spectrum becomes smaller if some fraction of the ionizing photons escape directly into the IGM, information about the LyC escape fraction is imprinted in the non-ionizing ($\lambda> 912$ \AA{}) part of the SED. This is demonstrated in Fig.~\ref{spectra}, where we present model SEDs for young starburst galaxies (age 10 Myr) at metallicities $Z=0.020$ (Fig.~\ref{spectra}a) and $Z=0.0004$ (Fig.~\ref{spectra}b). The line colours in each panel represent different $f_\mathrm{esc}$ (0.0, 0.3, 0.5 and 0.7). At both metallicities, lower $f_\mathrm{esc}$ imply stronger emission lines and a more prominent Balmer jump at 3646 \AA{}. In Fig.~\ref{spectra}b, the entire slope of the UV continuum changes with $f_\mathrm{esc}$, beacuse of the higher relative contribution from nebular emission at low metallicity. Similar changes in the UV slope are seen at higher metallicities as well, albeit at younger starburst ages. The noticeable dependence of several spectral features on $f_\mathrm{esc}$ suggest that it should be possible to assess the LyC escape fraction from galaxy SEDs without actually measuring the LyC flux. The question is simply how to best retrieve the $f_\mathrm{esc}$ information from the rest-frame UV/optical SED that can actually be observed in the reionization epoch ($\lambda> 1216$ \AA{}, since the flux at shorter wavelengths is absorbed by the neutral IGM).

\subsection{The UV slope $\beta$}
\begin{figure*}
\plottwo{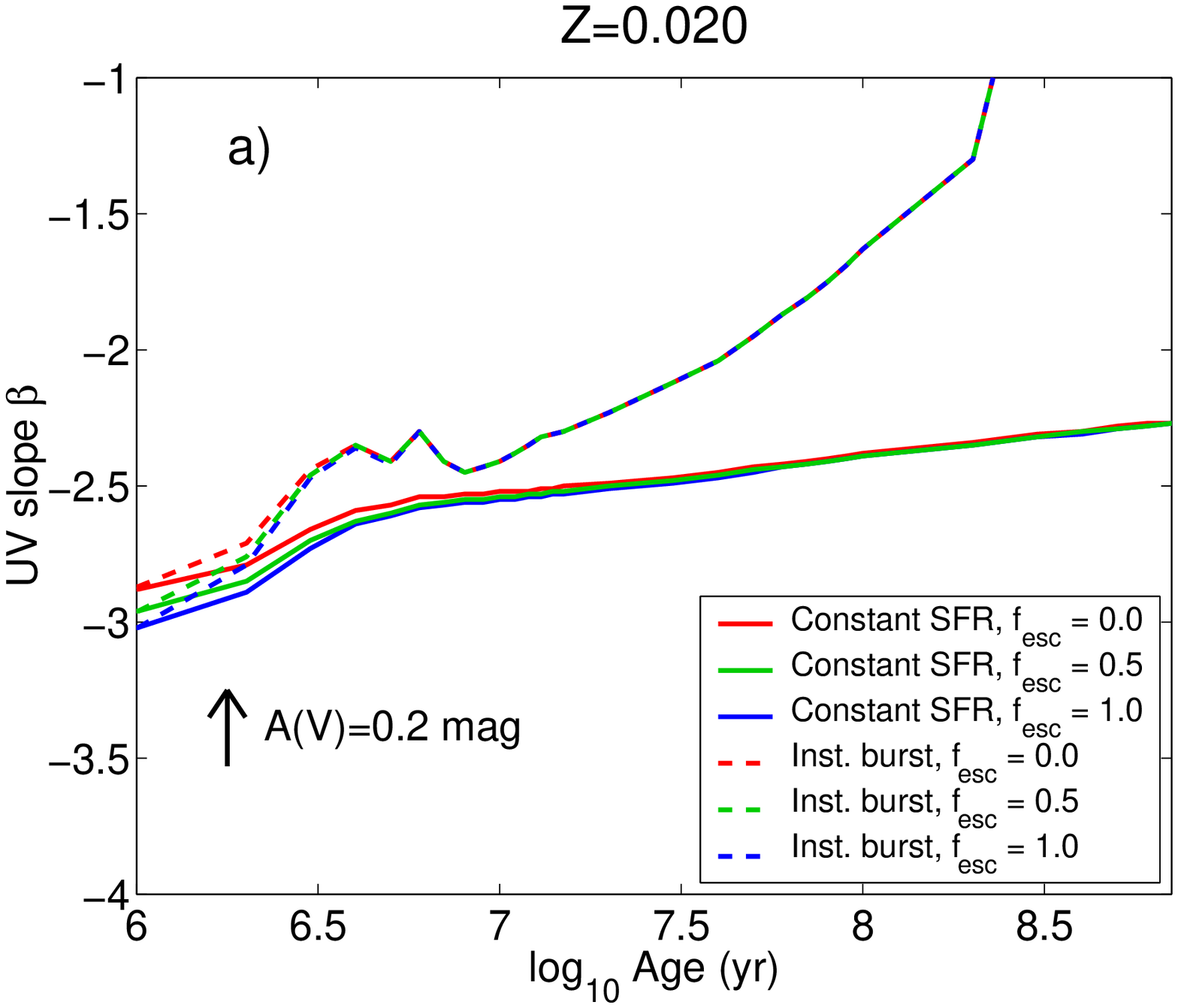}{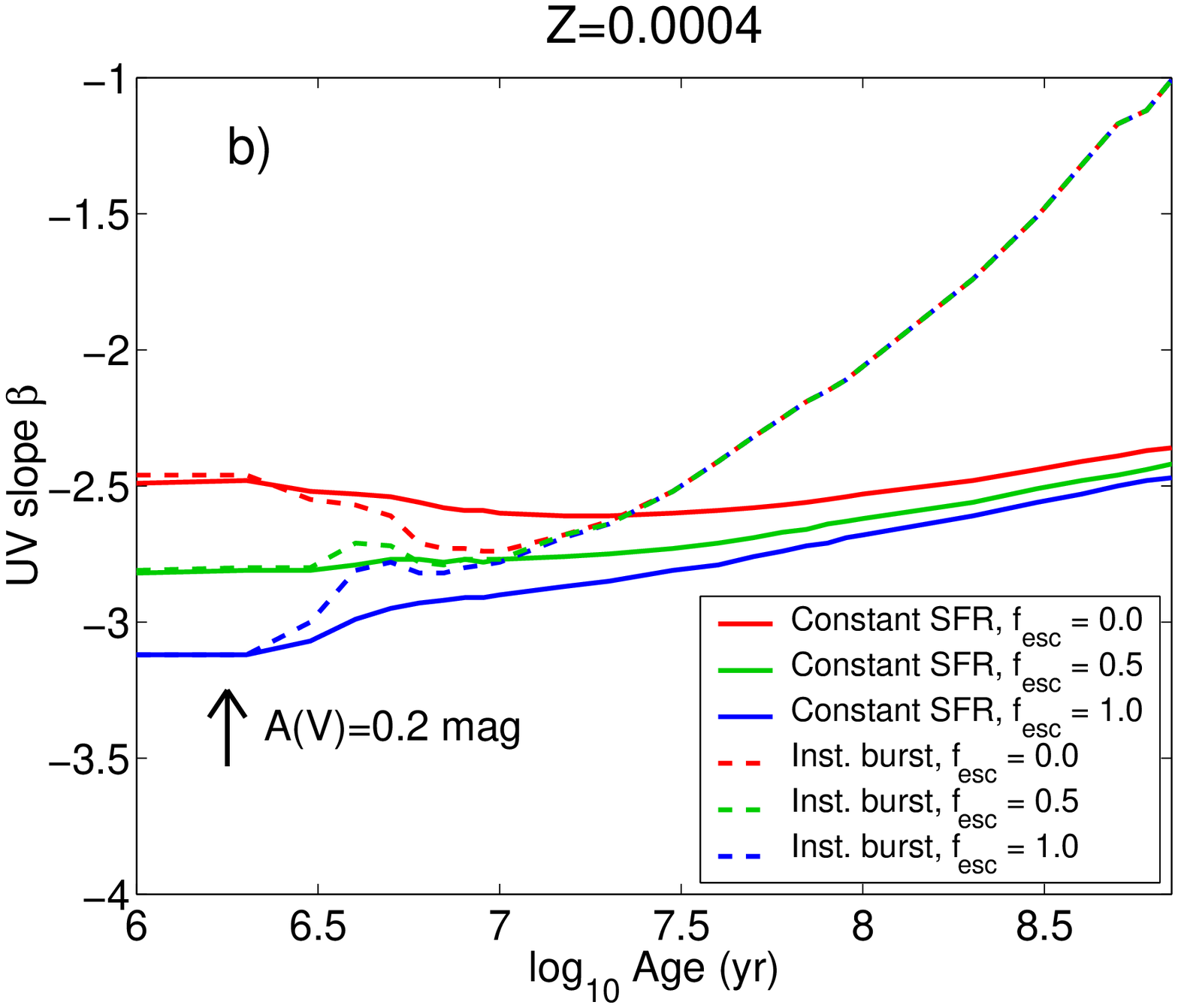}
\caption{The predicted evolution of the UV slope $\beta$ as a function of age for young starbursts with different LyC escape fractions at {\bf a)} $Z=0.020$ and {\bf b)} $Z=0.0004$. The different line colours represent $f_\mathrm{esc}=0.0$ (red), 0.5 (green) and 1.0 (blue). Solid lines indicate models with constant star formation rates and dashed lines models with instantaneous bursts. The arrows show how $A(\mathrm{V})$=0.2 mag of \citet{Calzetti00} extinction would shift $\beta$ in the case of $f_\mathrm{esc}=0.0$ (the effect is smaller at lower $f_\mathrm{esc}$, since we here assume the geometry depicted in Fig.~\ref{geometries}a, where the star light that escapes through holes in the nebula is unaffected by dust).\label{UV_slope}}
\end{figure*}
The power-law slope $\beta$ ($f_\lambda\propto \lambda^\beta$) of the UV continuum $f_\lambda$ can in principle be used to gauge $f_\mathrm{esc}$, since the addition of nebular emission to a young, purely stellar SED tends to shift the slope of the UV continuum in the redward direction and increase $\beta$ \citep[e.g.][]{Raiter et al.,Robertson et al. a}. The interpretation becomes ambiguous, however, unless a very blue UV slope is detected ($\beta \lesssim -3$). The distribution of $\beta$ slopes among galaxies in the reionization epoch is still a matter of debate. \citet{Bouwens et al. a} reported an average $\overline{\beta}\approx -3$ for $z\approx 7$, but later studies have given no support for a typical slope this extreme and instead indicate $\overline{\beta}\approx -2$ \citep[e.g.][]{Schaerer & de Barros b,McLure et al.,Bouwens et al. c,Finkelstein et al. a,Dunlop et al.,Rogers et al.}. It is possible, however, that a subset of objects still display UV slopes substantially bluer than this \citep[e.g.][]{Finkelstein et al. a,Jiang et al.}. 

The problem in inferring $f_\mathrm{esc}$ from $\beta$ is that $\beta$ also depends on age, metallicity and dust reddening \citep[e.g.][]{Schaerer & Pello,Bouwens et al. a}. This is demonstrated in Fig.~\ref{UV_slope}, where we plot the age dependence of $\beta$ for $Z=0.020$ (Fig.~\ref{UV_slope}a) and $Z=0.0004$ (Fig.~\ref{UV_slope}b) galaxies with either single-age stellar populations (dashed lines) or constant star formation rates (solid lines). Different line colours are used to indicate the impact of the LyC escape fraction in the case of an radiation-bounded nebula with holes (Fig.~\ref{geometries}a). 

The arrow shows the shift in $\beta$ ($\Delta(\beta)\approx 0.2$) predicted for $A(V)=0.2$ mag of $V$-band extinction of the stellar continuum (corresponds to $\approx 0.6$ mag at a rest wavelength of 1500 \AA{}) in the case of $f_\mathrm{esc}=0$, assuming the \citet{Cartledge et al.} average Large Magellanic Cloud (LMC) extinction curve. This amount of UV/optical extinction is consistent  with several current estimates of the amount of dust affecting the UV slopes of $z\gtrsim 7$ galaxies \citep[e.g.][]{Finkelstein et al. a,Dunlop et al.,Bouwens et al. c,Wilkins et al.}. At fixed $A(V)=0.2$ mag, the \citet{Calzetti00} attenuation curve gives a similar result ($\Delta(\beta)\approx 0.2$). As seen in Fig.~\ref{UV_slope}, the effect of $f_\mathrm{esc}$ on $\beta$ is very small for metal-rich galaxies (Fig.~\ref{UV_slope}a) but becomes substantial at low metallicities (Fig.~\ref{UV_slope}b) due to greater importance of nebular emission in the latter case. But even at metallicities as low as $Z=0.0004$, it is only for extremely blue UV slopes ($\beta\leq -2.7$) that any sort of constraint on $f_\mathrm{esc}$ can be set from $\beta$ alone. 

Here, we use the original definition of $\beta$ given by \citet{Calzetti et al.}, in which the UV slope is derived from the overall continuum flux (stellar and nebular) in 10 wavelength intervals (chosen to avoid stellar and interstellar absorption features) in the range $\approx 1270$--2580 \AA{}. The exact value of $\beta$ at any given age depends a bit on the exact wavelength region over which this slope is measured \citep{Raiter et al.}, but the overall trends seen in Fig.~\ref{UV_slope} remain the same, regardless of the definition of $\beta$ used. It should be noted, however, that the 2175 \AA{} dust feature can interfere with a few of the \citet{Calzetti et al.} bands and make the UV slope behave in unexpected ways. In this paper, we therefore refrain from discussing attenuation recipies with very prominent 2175 \AA{} bumps, like the average Milky Way extinction curve \citep[e.g.][]{Gordon et al.} . 
 
\subsection{The EW(H$\beta$)-$\beta$ diagram}
\label{EW_beta}
\begin{figure}
\plotone{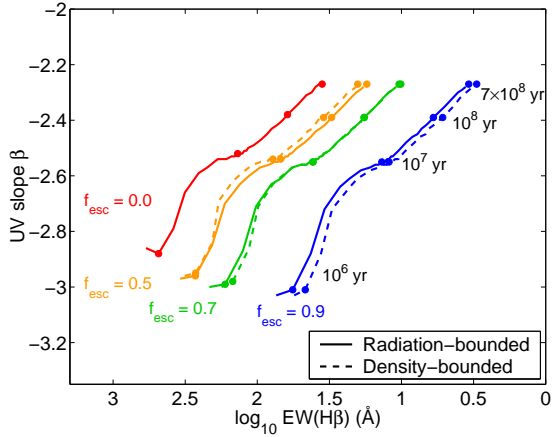}
\caption{EW(H$\beta$) versus $\beta$ at $Z=0.020$. A constant SFR is assumed for all models. Ages of 1, 10, 100 and 700 Myr (roughly the age of the Universe at $z=7$) are marked by filled circles along each track. Solid lines indicate radiation-bounded models with holes (Fig.~\ref{geometries}a) and dashed lines density-bounded models (Fig.~\ref{geometries}b). The different line colours represent different LyC escape fractions: $f_\mathrm{esc}=0$ (red),  $f_\mathrm{esc}=0.5$ (orange),  $f_\mathrm{esc}=0.7$ (green) and  $f_\mathrm{esc}=0.9$ (blue). The fact that these tracks remain disconnected up to the highest ages considered indicates that it should be possible to assess $f_\mathrm{esc}$ from a simulataneous measurement of EW(H$\beta$) and $\beta$. 
\label{HB_beta_geom}}
\end{figure}

We suggest that by combining the UV slope $\beta$ with the equivalent width (EW) of a Balmer line such as H$\beta$, it should be possible to identify galaxies with high escape fractions ($f_\mathrm{esc}\geq 0.5$) as long as the UV slope is $\beta \leq -2.3$ after dust reddening corrections (not far from the typical slope measured at $z\gtrsim 6$, which is likely to be at least slightly affected by dust; \citealt{McLure et al.,Bouwens et al. c,Finkelstein et al. a,Dunlop et al.,Wilkins et al.}). This idea is demonstrated in Fig.~\ref{HB_beta_geom}, where we plot the rest-frame EW(H$\beta$) against $\beta$ for $Z=0.020$ for constant-SFR, dust-free models with different LyC escape fractions. Results are shown for both geometries depicted in Fig.~\ref{geometries}, i.e. an radiation-bounded nebula with holes (solid lines) and a density-bounded nebula (dashed lines). 

The high-leakage models ($f_\mathrm{esc}\geq 0.5$; orange, green and blue lines) are all shifted to the right in Fig.~\ref{HB_beta_geom} (i.e. towards lower EW(H$\beta$)) compared to the no-leakage case (red lines). Since the model tracks remain separated up to the highest age considered (700 Myr; approximately the age of the Universe at $z=7$), this diagram suggests that $f_\mathrm{esc}$ can be assessed from a simultaneous measurement of EW(H$\beta$) and $\beta$ (additional diagnostics can however improve the quality of this estimate, as discussed below). The results for radiation-bounded and the density-bounded models are next to identical at $f_\mathrm{esc}>0$ (and, by definition, completely identical for $f_\mathrm{esc}=0$). In the following, we will therefore focus on the radiation-bounded models.

At this metallicity ($Z=0.020$), $\beta$ is seen to change very little as $f_\mathrm{esc}$ is increased, simply due to the fact that nebular emission has such a small impact on the UV slope (Fig.~\ref{UV_slope}). The equivalent width EW(H$\beta$) of course drops with increasing $f_\mathrm{esc}$, but since nebular emission produces both the H$\beta$ emission line and part of the continuum below the line, the decrease is not as dramatic as one would expect if one assumed the underlying continuum to be purely stellar. Similar diagnostic diagrams can in principle be formed for other Balmer lines (H$\alpha$, H$\gamma$, H$\delta$ etc.), but H$\beta$ seems to be a good compromise between observational limitations (H$\alpha$ redshifts out of the JWST/NIRSpec range already at $z\geq 6.6$) and expected diagnostic value (Balmer lines beyond $H\beta$ are weaker and have lower equivalent widths). Other definitions of $\beta$ give rise to slightly different diagnostic diagrams, but the overall trends remain the same. While the [OIII]$\lambda$5007 emission line can be measured over approximately the same redshift interval as H$\beta$, and will in many cases be stronger, the equivalent width of this line is not as suitable as a diagnostic, since it is very sensitive to the ionization parameter of the gas.  

\subsection{Metallicity}
\label{metallicity}
\begin{figure}
\plotone{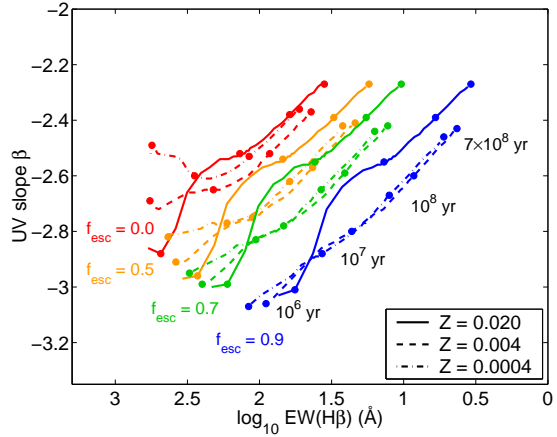}
\caption{EW(H$\beta$) versus $\beta$ for galaxies with metallicity $Z=0.020$ (solid lines), $Z=0.004$ (dashed) and $Z=0.0004$ (dash-dotted). The line colours represent different LyC escape fractions: $f_\mathrm{esc}=0$ (red),  $f_\mathrm{esc}=0.5$ (orange),  $f_\mathrm{esc}=0.7$ (green) and  $f_\mathrm{esc}=0.9$ (blue). A constant SFR is assumed for all models. Ages of 1, 10, 100 and 700 Myr (roughly the age of the Universe at $z=7$) are marked by filled circles along each track. Numerical age labels are also given for the $Z=0.0004$ track. The fact that lines of different colour sometimes cross each other at ages lower than 700 Myr (most notably orange and green lines) means that there will be some uncertainty on the inferred $f_\mathrm{esc}$ unless the metallicity is estimated through other means, e.g. the [NeIII]$\lambda$3869, [OII]$\lambda$3727 and [OIII]$\lambda$4959, 5007 emission lines. 
\label{HB_beta_Z}}
\end{figure}

Since metallicity is an important parameter in regulating the impact of nebular emission in galaxies \citep[][]{Zackrisson et al. a}, one expects diagnostic diagrams such as Fig.~\ref{HB_beta_geom} to show some metallicity dependence. In Fig.~\ref{HB_beta_Z}, we display the same $Z=0.020$, constant-SFR model tracks for radiation-bounded nebulae as in Fig.~\ref{HB_beta_geom}, alongside otherwise identical models at $Z=0.004$ and $Z=0.0004$. As expected, $\beta$ evolves more strongly with $f_\mathrm{esc}$ at low metallicities (see Fig.~\ref{UV_slope}), whereas EW(H$\beta$) displays a milder metallicity dependence. The latter effect is due to the greater contribution from nebular emission to {\it both} the H$\beta$ emission line and the underlying continuum at low metallicities.  While a rough assessment of $f_\mathrm{esc}$ can be obtained even if the metallicity is unknown, a more precise estimate requires some handle on the metallicity. This is most readily obtained from the strengths of emission lines like [NeIII]$\lambda$3869, [OII]$\lambda$3727 and [OIII]$\lambda$4959, 5007 \citep[e.g.][]{Nagao et al.}, since data on these lines come ``for free'' when doing spectroscopy over the rest-frame wavelength interval ($\approx 1200$--5000 \AA{}) required to measure $\beta$ and EW(H$\beta$). It should be noted, however, that many of the metallicity calibrations developed for these lines are based on the assumption of no LyC leakage, and may potentially give biased results if blindly applied to density-bounded nebulae with high $f_\mathrm{esc}$  \citep[e.g.][]{Giammanco et al.}.

\subsection{Star formation history}
\begin{figure*}
\plottwo{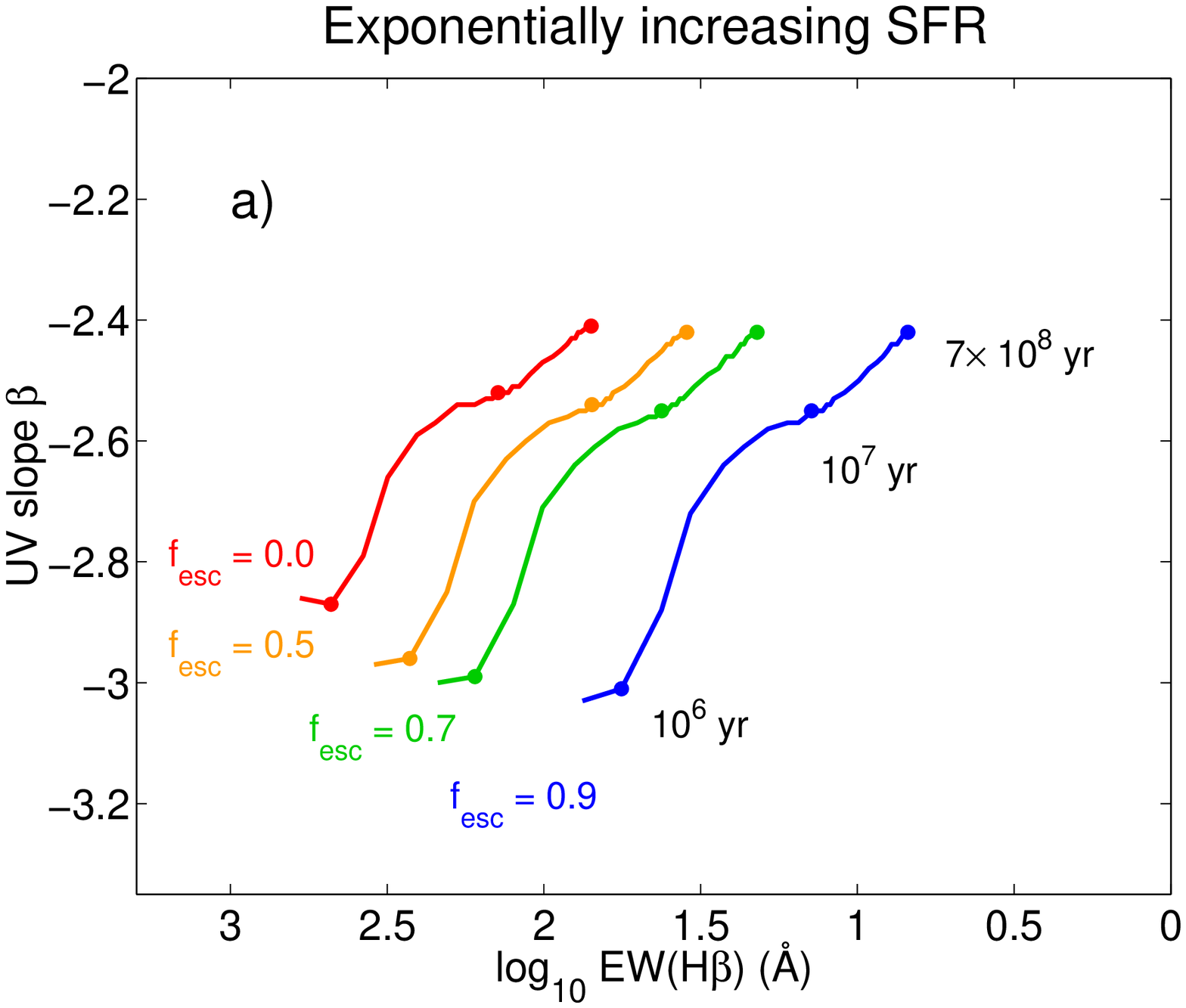}{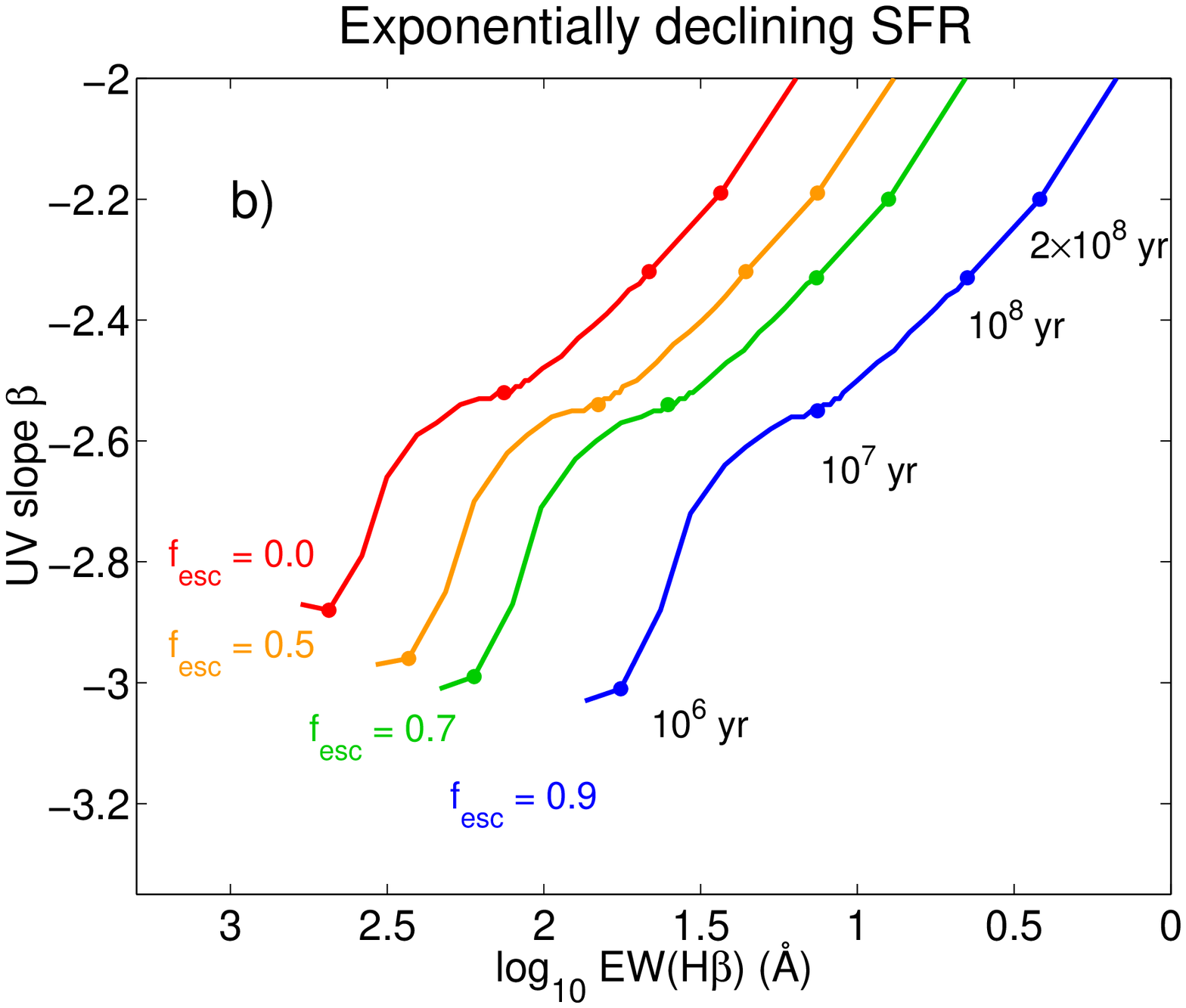}
\plottwo{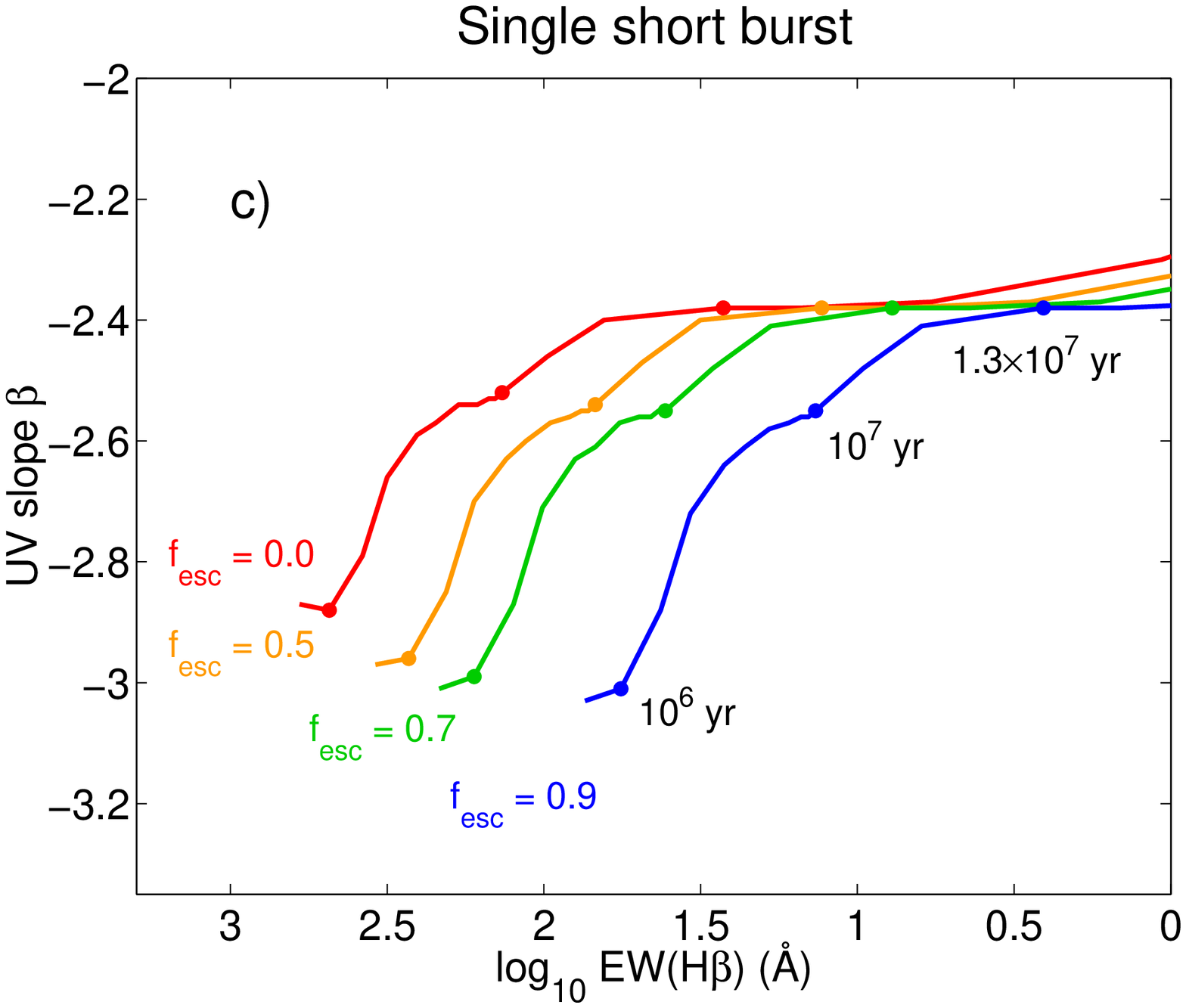}{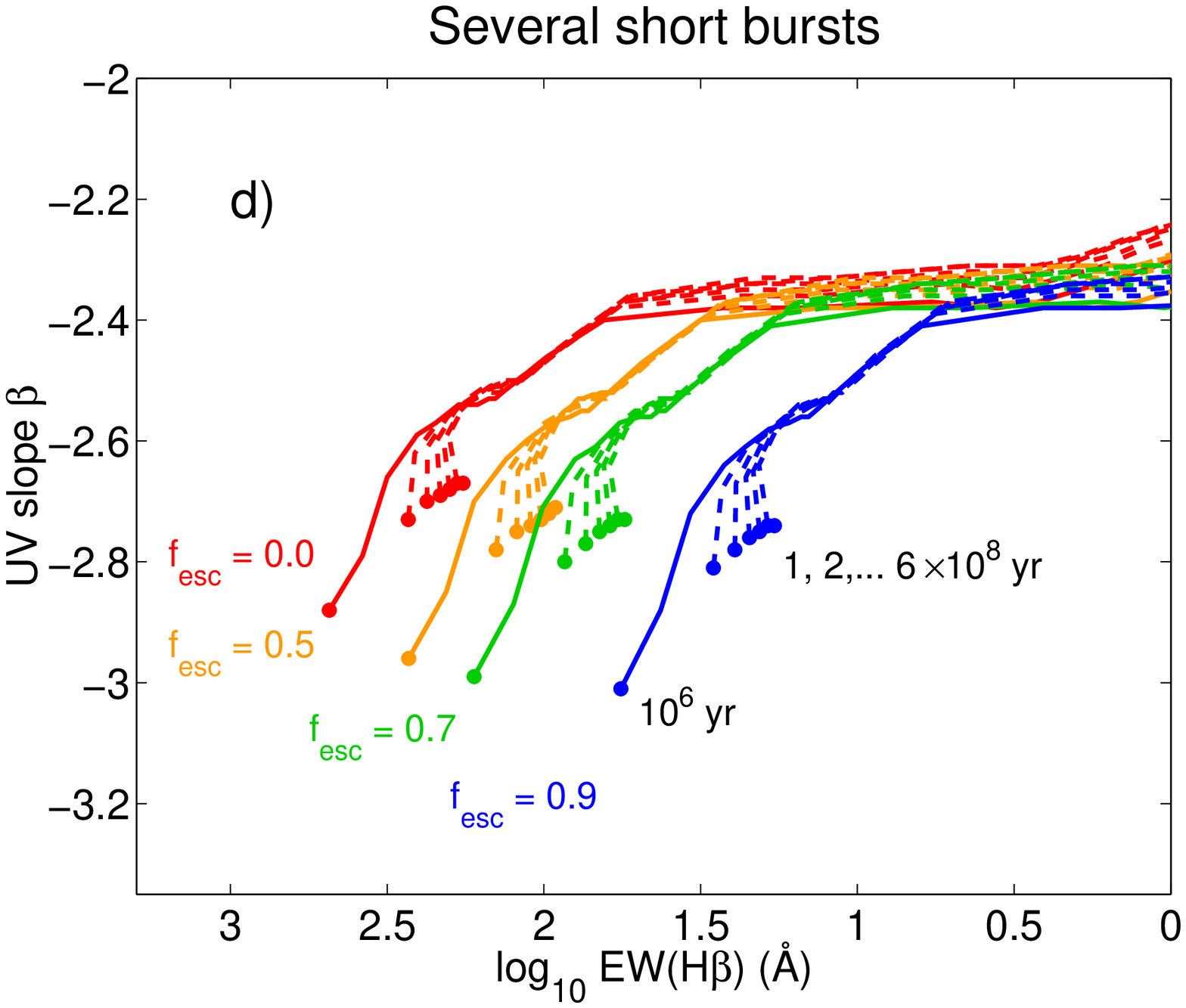}
\caption{The effect of star formation history variations on the EW(H$\beta$)-$\beta$ diagnostic. All models assume $Z=0.020$, an radiation-bounded nebula with holes but different star formation histories: {\bf a)} An exponentially increasing star formation rate ($\mathrm{SFR}(t)\propto \exp(-t/\tau)$ with $\tau=-10^8$ yr); {\bf b)} An exponentially decreasing star formation rate ($\tau=10^8$ yr); {\bf c)} A short burst of constant SFR (lasting $10^7$ yr), after which the SFR drops to zero; {\bf d)} A sequence of short, constant-SFR bursts (each lasting $10^7$ yr) every $10^8$ yr with no SFR in between. The line colours represent different LyC escape fractions: $f_\mathrm{esc}=0$ (red),  $f_\mathrm{esc}=0.5$ (orange),  $f_\mathrm{esc}=0.7$ (green) and  $f_\mathrm{esc}=0.9$ (blue). Only in the short burst scenario is there (at $\beta \approx -2.4$) any room for serious confusion concerning $f_\mathrm{esc}$, but the time spent in this part of the diagram is very brief ($\approx 2$ Myr), making misclassifications unlikely.
\label{HB_beta_SFH}}
\end{figure*}
\begin{figure*}
\plottwo{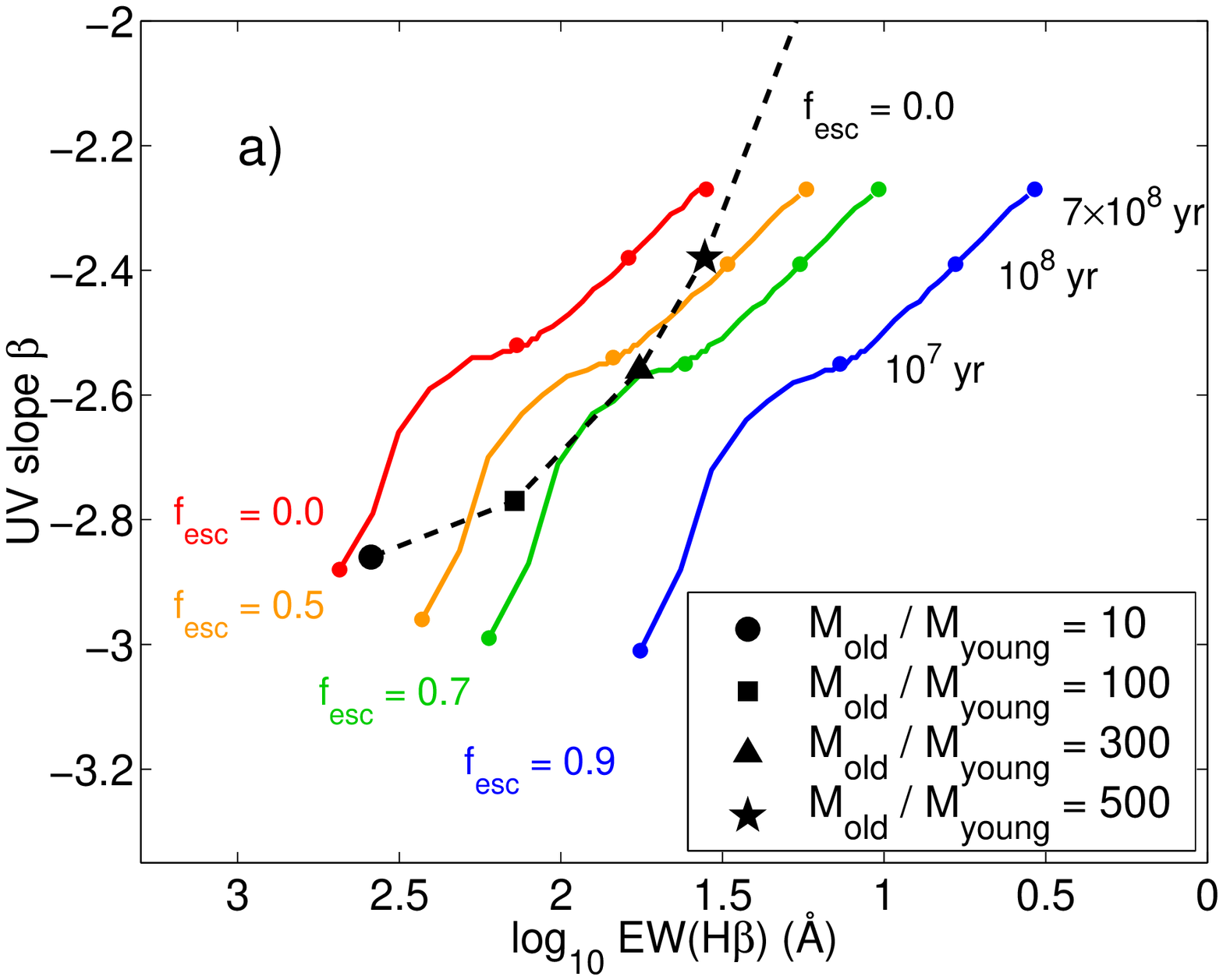}{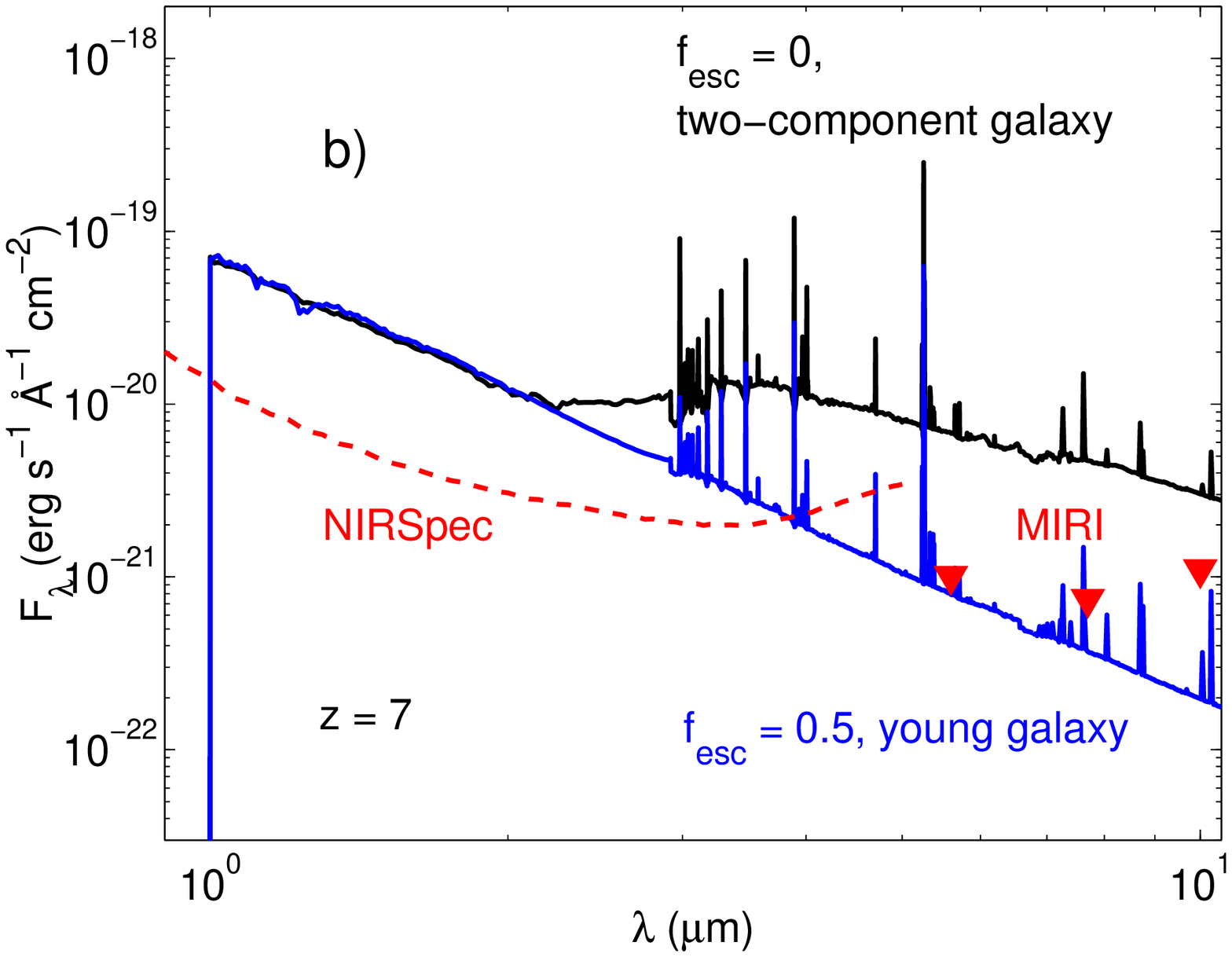}
\caption{Examples of how young bursts in old galaxies can mimic LyC leakage in the EW(H$\beta$) vs. $\beta$ diagram. {\bf a)} Constant-SFR, radiation-bounded models with $Z=0.020$ and different $f_\mathrm{esc}$ (red, orange, green and blue lines) compared to a very young ($10^6$ yr old) burst with $f_\mathrm{esc}=0$ superposed on a $7\times 10^8$ yr old, passively evolving stellar population with different mass ratios $M_\mathrm{old}/M_\mathrm{young}$ (black dashed line with markers at $M_\mathrm{old}/M_\mathrm{young}=10$, 100, 300 and 500). Since the old population contributes substantially to the H$\beta$ continuum, but less to the UV slope $\beta$, such compound populations can in principle mimic genuinely young bursts with high $f_\mathrm{esc}$ in this diagram. Age markers along the constant-SFR tracks represent ages of $10^6$, $10^7$ and $10^8$ and $7\times 10^8$ yr. {\bf b)} The $z=7$ SED of the $M_\mathrm{old}/M_\mathrm{young}=300$ compound population (black line) from the left panel (black triangle) compared to the $z=7$ SED of a genuinely young ($10^7$ yr old), $f_\mathrm{esc}=0.5$ burst (blue line). The stellar mass is $4\times 10^{10}\ M_\odot$ for the compound object and $2\times 10^8\ M_\odot$ for the young galaxy. While having nearly identical $\beta$ and EW(H$\beta$), the compound object has a much higher continuum flux at $\lambda\gtrsim 3\ \mu$m, which means that a JWST/NIRSpec spectroscopy or JWST/MIRI imaging should be able to break the $f_\mathrm{esc}$ degeneracy. The 10$\sigma$ detection limits for low-resolution ($R=100$) spectroscopy with JWST/NIRSpec after a 10 h exposure are indicated by the red dashed line. The corresponding 10$\sigma$ imaging detection limits after a 10 h exposure are shown for the three JWST/MIRI filters at $\lambda\leq 10 \ \mu$m are shown by red triangles. 
\label{HB_beta_SFH2}}
\end{figure*}

As illustrated in Fig.~\ref{HB_beta_SFH}, temporal variations of the star formation rate (SFR) can affect both EW(H$\beta$) and $\beta$ and need to be considered when attempting to assess the LyC escape fraction. Galaxies with increasing star formation rates simply remain longer in the high EW(H$\beta$), blue $\beta$ part of the EW(H$\beta$)-$\beta$ diagram, which makes the determination of $f_\mathrm{esc}$ easier than in the case of a constant SFR (Fig.~\ref{HB_beta_geom}). This is shown in Fig.~\ref{HB_beta_SFH}a, where $Z=0.020$ models with exponentially increasing SFRs ($\mathrm{SFR}(t)\propto \exp(-t/\tau)$ with $\tau=-10^8$ yr) at various $f_\mathrm{esc}$ are used. Models with decreasing star formation rates are, however, more troublesome. In Fig.~\ref{HB_beta_SFH}b, we show the behaviour of models with exponentially declining star formation rates ($\tau=10^8$ yr). These models evolve faster than the constant-SFR models of Fig.~\ref{HB_beta_geom} and reach slightly redder $\beta$ at the highest ages considered. However, as there is no overlap between the different $f_\mathrm{esc}$ tracks, it should still be possible to estimate the LyC escape fraction. In Fig.~\ref{HB_beta_SFH}c, we consider a more extreme model with a very short burst ($10^7$ yr) of constant star formation, after which the SFR immediately drops to zero. Since the Lyman continuum flux decreases very quickly once star formation has ceased, models of this type evolve sharply to the right in the EW(H$\beta$)-$\beta$ diagram once the burst is over. During the post-burst phase, there is a brief period where model lines corresponding to different $f_\mathrm{esc}$ overlap, which means that an accurate assessment of $f_\mathrm{esc}$ would be very difficult in this part of the diagram. For these particular models, this happens at ages $\approx 13$--15 Myr ($\beta\approx -2.4$), when the LyC flux has dropped to 2--15\% of what it was right before star formation ceased. We stress, however, that galaxies will only rarely be caught in this brief transition phase.

Single-component models of the type used in Fig.~\ref{HB_beta_SFH}a--c are often assumed to be adequate in the interpretation of observational data on high-redshift galaxies, but real galaxies may well exhibit more complicated star formation histories, with several bursts with intermittent periods of low star formation in between. In Fig.~\ref{HB_beta_SFH}d, we consider a sequence of equal-strength, $10^7$ yr long, constant-SFR bursts once every $10^8$ yr, with zero SFR in between. Each burst shifts the models to slightly redder $\beta$ and lower EW(H$\beta$), thereby causing a slight overlap between models with different $f_\mathrm{esc}$. 

In this context, the worst-case scenario would be to have a single young, low-$f_\mathrm{esc}$ burst superposed on an old, passively evolving population which contributes substantially to the H$\beta$ continuum (thereby lowering EW(H$\beta$)) while leaving the UV slope $\beta$ largely unaffected, since this could in principle mimic high $f_\mathrm{esc}$ in the EW(H$\beta$)-$\beta$ diagram. In Fig.~\ref{HB_beta_SFH2}a, we show the evolution of such a composite population, where a 1 Myr old, $f_\mathrm{esc}=0$  model has been superposed on a 700 Myr old, passively evolving population. As the mass ratio between the old population and the short burst is increased to $\gtrsim 30$, this $f_\mathrm{esc}=0$ model can wander into the region normally occupied by leaking galaxies with $f_\mathrm{esc}\gtrsim 0.5$. If the mass ratio is increased to $>500$, the passively evolving population becomes completely dominant also in the UV, and shifts $\beta$ to very red values. However, the presence of an old, massive component gives rise to other tell-tale spectral signatures at the wavelengths probed by both NIRSpec (0.6--5 $\mu$m) and the MId-Infrared Instrument (MIRI; 5--27 $\mu$m in imaging mode) onboard the JWST. 

This is demonstrated in Fig.~\ref{HB_beta_SFH2}b, where we plot the SEDs of a 10 Myr old, constant-SFR model with $f_\mathrm{esc}=0.5$ (blue line) and a 1 Myr old $f_\mathrm{esc}=0$ model superposed on a passively evolving population with age 700 Myr (black line), where the old component is 300 times massive than the young population. Both spectra have been redshifted to $z=7$. These two models occupy a very similar position in the EW(H$\beta$)-$\beta$ diagram (black triangle and orange $10^7$-yr age marker in Fig.~\ref{HB_beta_SFH2}a), but the continuum flux observed at $\gtrsim 3\mu$m would be very different in the two cases (much higher in the case of the two-component galaxy). 

The continuum flux at $\leq 5\ \mu$m comes for free when measuring $\beta$ and EW(H$\beta$) with JWST/NIRSpec, and JWST/MIRI can trace the continuum to even longer wavelengths. MIRI lacks the sensitivity to provide useful spectroscopic limits on the continuum, but MIRI imaging should yield competitive constraints. This is demonstrated in Fig.~\ref{HB_beta_SFH2}b, where we include the expected detection limits for JWST/NIRSpec low-resolution ($R=100$), 1--5 $\mu$m spectroscopy  (red dashed line) and compare these to the corresponding limits for JWT/MIRI imaging (red triangles) at $\approx 5$--10 $\mu$m. In both cases, we adopt an exposure time of 10h and require a $S/N=10$ detection.

\subsection{Dust attenuation}
\label{attenuation}
In the quest to derive $f_\mathrm{esc}$ from the rest-frame UV/optical SED, dust attenuation is potentially a more serious problem than metallicity or star formation history, since this part of the SED may contain insufficient information to gauge the impact of LyC extinction. Many recent studies of reionization-epoch galaxies suggest that these objects have suffered very little dust attenuation ($A(\mathrm{V})\lesssim 0.2$ mag; e.g. \citealt{Finkelstein et al. a,Bouwens et al. c,Dunlop et al.,Wilkins et al.}, but see \citealt{Schaerer & de Barros b} for a different view). On the other hand, it doesn't necessarily take much attenuation in the rest-frame optical or non-ionizing UV to have a considerable impact on the LyC -- it depends on the spatial distribution of the dust compared to the gas and stars. 
\begin{figure*}
\plotone{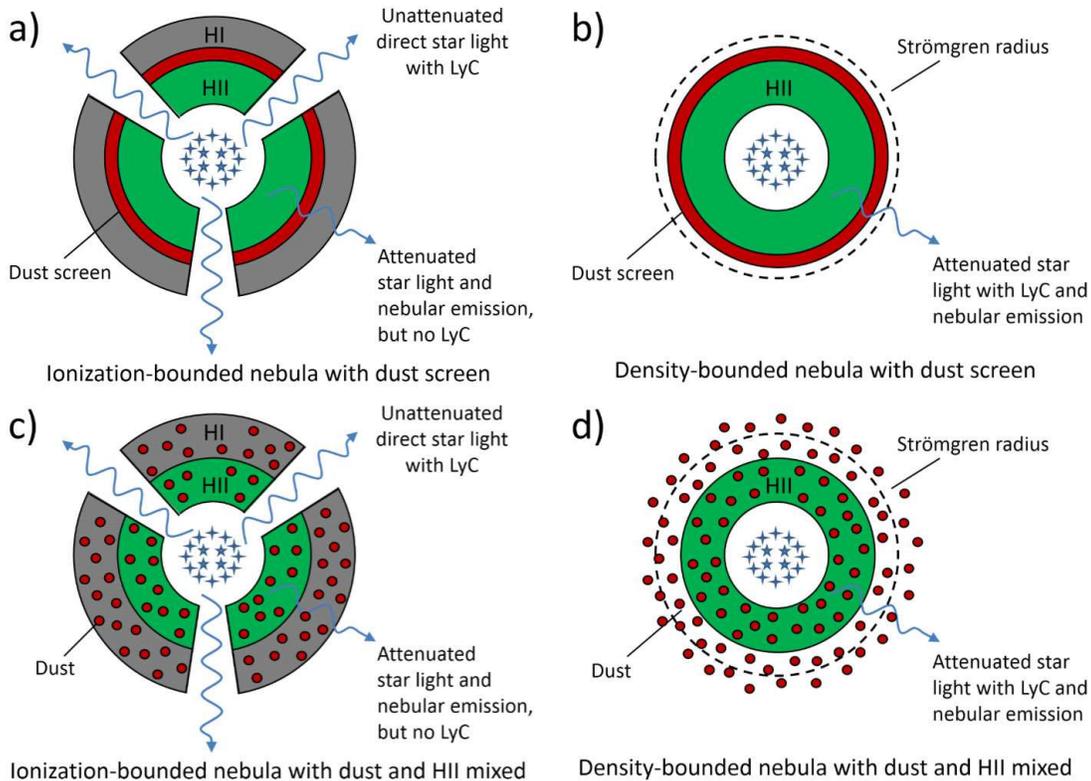}
\caption{Different dust distributions: {\bf a)} radiation-bounded nebula with an outer dust screen (both with holes); {\bf b)} density-bounded nebula with an outer dust screen; {\bf c)} radiation-bounded nebula with holes and dust mixed with the ionized gas; {\bf d)} Density-bounded nebula with dust mixed with the ionized gas.
\label{dust_geometry}}
\end{figure*}

\subsubsection{Dust distributions}
Predicting the effects of dust on galaxy SEDs is notoriously difficult, since the effective dust attenuation law depends not only on the amount and composition of dust, but also on the spatial distribution of stars dust and gas \citep[e.g.][]{Calzetti et al.,Gordon et al. 97,Witt & Gordon}. In Fig.~\ref{dust_geometry}, we schematically illustrate four different dust distributions (two for each of the gas geometries of Fig.~\ref{geometries}). This is not meant to be an exhaustive list of the possible dust distributions, but serves to illustrate a few different scenarios. In Fig.~\ref{dust_geometry}a and b, the dust is distributed in a uniform screen {\it outside} the HII region. In the cases depicted in Fig.~\ref{dust_geometry}a  and c (ioniziation-bounded nebulae with holes), dust has no direct impact on the leaking LyC flux, since the leakage is assumed to take place through holes that are devoid of both gas and dust. In the density-bounded nebula of Fig.~\ref{dust_geometry}b, on the other hand, the dust screen has a very pronounced effect on the escaping LyC. As an example of this, consider a dust screen giving a rest-frame $V$-band extinction of $A(\mathrm{V})=0.2$ mag. Extinction curves are notoriously uncertain in the far-UV, but if one adopts the average LMC attenuation curve presented by \citet{Cartledge et al.}, this $A(V)$ converts into a predicted LyC depletion factor of $\approx 3$ at 910 \AA{}\footnote{This is admittedly an extrapolation, since the \citet{Cartledge et al.} analysis is based on data at $\geq 1050$ \AA{}, but the \citet{Gordon et al.} Milky Way extinction curve -- which extends to shorter wavelengths -- predicts a similar LyC depletion factor}. Hence, the LyC escape fraction would effectively be limited to $f_\mathrm{esc} < 1/3$.

In the case of a foreground dust screen (Fig.~\ref{dust_geometry}a and b), the rest-frame optical attenuation can straightforwardly be estimated from JWST/NIRSpec spectra of $z\approx 6$--9 galaxies by measuring of the ratios of Balmer emission-lines (e.g. H$\beta$/H$\gamma$), since the intrinsic ratio can be estimated from recombination theory (H$\beta$/H$\gamma\approx 2.1$--2.2 for a wide range of electron densities and temperatures, assuming case B recombination). However, if the dust is spatially mixed with the HII gas as in Fig.~\ref{dust_geometry}c and d, some fraction of the LyC photons will be absorbed by dust before they have a chance to ionize the gas \citep[e.g.][]{Inoue et al. a,Inoue a}, and this will not be revealed by the Balmer line ratios. While radiation pressure and dust sublimation act to produce a dust cavity around the ionizing stars, neither of these effects is sufficiently strong to prevent LyC photons from getting absorbed directly by dust in local HII regions \citep{Inoue b}. Indeed, \citet{Hirashita et al.} estimate that $\approx 40\%$ of the LyC photons produced in low-redshift starburst galaxies are typically absorbed by dust. 

\subsubsection{Dust effects in the EW(H$\beta$)-$\beta$ diagram}
\begin{figure}
\plotone{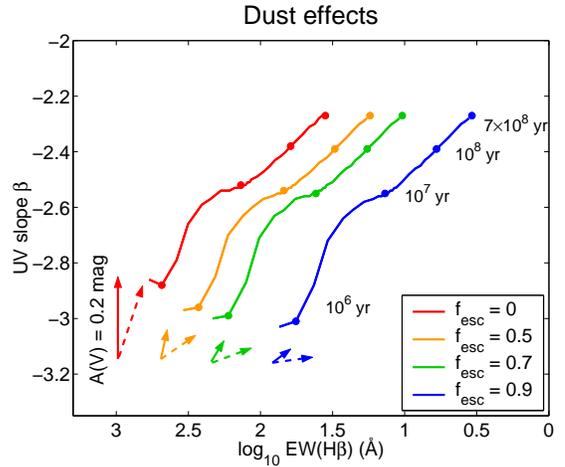}
\caption{EW(H$\beta$) versus $\beta$ at $Z=0.020$ for constant-SFR models, with attenuation vectors for different dust distributions. Ages of 1, 10, 100 and 700 Myr are indicated by markers along each track. The line colours represent different $f_\mathrm{esc}$. The arrows show how dust attenuation is likely to shift the position of radiation-bounded galaxies with holes in this diagram, under the assumption of $A(V)=0.2$ mag and an average LMC attenuation law \citep{Cartledge et al.}. The solid arrows represent the shift in position produce by an outer dust screen with holes (Fig.~\ref{dust_geometry}a), whereas the dashed arrows indicates the shift when dust is spatially mixed with gas (Fig.~\ref{dust_geometry}c), and the nebular photons only experiences half of the extinction suffered by the stars.  
\label{dust_corrections}}
\end{figure}

Since simulations of high-redshift galaxies suggest that when LyC escapes, it does so through essentially dust-free channels \citep{Gnedin et al.,Razoumov & Sommer-Larsen}, we will limit the discussion to radiation-bounded nebulae with holes (Fig.~\ref{dust_geometry}a and c). For both of these geometries, the effects of small amounts of optical attenuation ($A(\mathrm{V})=0.2$ mag) on positions of galaxies in the EW(H$\beta$)-$\beta$ diagram are illustrated in Fig..~\ref{dust_corrections}, assuming the \citet{Cartledge et al.} average LMC attenuation law to capture the transport of radiation through dusty regions. 

At low $f_\mathrm{esc}$, a foreground dust screen (attenuation vector represented by solid arrow) has a pronounced effect on $\beta$ but almost no effect on EW(H$\beta$), since the stellar continuum emerging through the screen still dominates the rest-frame UV, while the unattenuated continuum that escapes through the holes remains subdominant at wavelengths close to H$\beta$. At high $f_\mathrm{esc}$, the reverse is true: $\beta$ is largely unaffected by dust since the unattenuated stellar UV continuum escaping through the holes dominates over the dust-attenuated continuum, and EW(H$\beta$) is lowered due to the increasing importance of stellar H$\beta$ continuum emerging through the holes. Hence, the attenuation vector changes direction and amplitude depending on where in the EW(H$\beta$)-$\beta$ diagram one started off. 

If the dust is spatially mixed with the ionized gas (Fig.~\ref{dust_geometry}c), the stellar non-ionizing continuum will still suffer $A(\mathrm{V})=0.2$ mag of attenuation, but the Balmer line ratios will reflect a value smaller than this, since part of the attenuation happened prior to photoionization. At fixed $f_\mathrm{esc}$, the nebula will also appear fainter due to LyC extinction. Here, we model this case by assuming that the star light entering the HII region is attenuated by $A(\mathrm{V})=0.1$ mag before it has a chance to ionize the gas (lowering the number of LyC photons that enter the gas by a factor of $\approx 2$), and apply a further $A(\mathrm{V})=0.1$ mag correction outside the HII region (giving a total of $A(\mathrm{V})=0.2$ mag, as before). Other ratios of stellar to nebular attenuation can certainly be considered, but this serves to exemplify the EW(H$\beta$) and $\beta$ trends. The resulting LyC depletion factor is somewhat lower than suggested by \citet{Inoue a} and \citet{Inoue et al. a} for local HII regions, but similar to that derived by \citet{Hirashita et al.} for low-redshift starburst galaxies. Our procedure gives rise to the dashed attenuation arrows in Fig.~\ref{dust_corrections}. As in the case of the outer dust screen, the $\beta$ slope is reddened more at low $f_\mathrm{esc}$ than at high $f_\mathrm{esc}$, but the overall reduction in nebular emission (due to LyC extinction) increases the relative impact of stellar H$\beta$ continuum on the emerging SED and decreases EW(H$\beta$) substantially at all $f_\mathrm{esc}$. 

At fixed $A(\mathrm{V})$, the two types of attenuation vectors discussed above have slightly different directions and amplitudes. Moreover, a measurement of the H$\beta$/H$\gamma$ Balmer decrement will tend to underestimate the overall attenuation whenever LyC photons are lost due to dust extinction, as in the case where ionized gas and dust are mixed. For a given H$\beta$/H$\gamma$ ratio, there will therefore be lingering uncertainties in how to correct the position in the EW(H$\beta$)-$\beta$ diagram unless the amount of LyC extinction can be assessed. For example, consider an object observed to lie close to the $10^7$ yr marker on the orange line ($f_\mathrm{esc}=0.5$) in Fig.~\ref{dust_corrections}. To correct the position of this object for dust effects, one needs to pick the right attenuation vector, and to move the object in the opposite direction in the diagram. If the Balmer line ratios suggest $A(\mathrm{V})=0.2$ mag, a correction assuming an outer dust screen would place the object on the green line (implying $f_\mathrm{esc}=0.7$), whereas a correction assuming mixed gas and dust would still place it on the orange line (i.e. $f_\mathrm{esc}=0.5$). These uncertainties are not easy to overcome from observations in the rest-frame UV/optical alone, but supporting observations in the rest-frame mid/far-IR could improve the situation (as further discussed in Sect.~\ref{IR}). 

The possibility that the effective attenuation law may have a more pronounced effect on emission lines than on the far-UV/optical continuum represents yet another complication. This phenomenon is observed in local starburst galaxies \citep[e.g.][]{Calzetti et al.,Calzetti00} and is most readily interpreted as due to temporal evolution in the opacity of star-forming regions \citep[e.g.][]{Charlot & Fall} -- the youngest star clusters (that contribute most of the LyC photons and the nebular emission) tend to be more strongly affected by dust than the slightly older star clusters, which contribute substantially to the far-UV/optical stellar continuum). In principle, this effect could mimic LyC leakage in the EW(H$\beta$)-$\beta$ diagram by causing a substantial reduction in EW(H$\beta$) but only a modest reddening of $\beta$. However, we find that the \citet{Calzetti00} attenuation scheme -- in which the nebular emission experiences a dust optical depth a factor of $\approx 2$ times than that of the stars -- results in an attenuation vector no more extreme than those already included in Fig.~\ref{dust_corrections}. In order for a $f_\mathrm{esc}=0$ galaxy to be misclassified as an $f_\mathrm{esc}\geq 0.5$ object, one requires both a higher ratio of nebular to stellar dust opacity -- as in some of the more extreme objects discussed by \citet{Wild et al.} -- and substantial nebular attenuation ($A(V)>0.3$ mag, as revealed by e.g. the H$\beta/$H$\gamma$ emission line ratio).  

\subsubsection{Mid/far-IR observations}
\label{IR}
The radiation absorbed by dust is reradiated in the rest-frame mid/far-IR, and combined observations of the rest-frame UV/optical and mid/far-IR have been used to estimate the fraction of LyC photons absorbed by dust in local objects \citep[e.g.][]{Inoue et al. a,Inoue a,Hirashita et al.}. This dust emission lies out of range of the JWST for $z\gtrsim 6$ objects, but as further discussed in Sect.~\ref{detection_limits}, observations with the Atacama Large Millimeter/submillimeter Array (ALMA) or the planned Space Infrared Telescope for Cosmology and Astrophysics (SPICA) may be able to probe dust emission from galaxies at these redshifts. There are, however, many outstanding problems involved. For instance, the dust is heated by both the LyC, the non-ionizing UV and the resonant Ly$\alpha$ line. Depending on the spatial distribution of gas, dust and stars and the overall velocity field, Ly$\alpha$ photons may escape from starburst galaxies at all redshifts, but for most galaxies in the reionization epoch, a substantial fraction of the Ly$\alpha$ photons that escape are scattered into huge and largely unobservable Ly$\alpha$ halos by the neutral IGM \citep[e.g.][]{Jeeson-Daniel et al.}, thereby causing potentially significant uncertainties in the IR radiation budget. The SED of dust emission also depends on the dust composition \citep[][]{Takeuchi et al.}, which is highly uncertain even for low-redshift targets.   

\section{Discussion}
\label{discussion}
\subsection{Detection limits}
\label{detection_limits}
The part of the SED required for the measurement of $\beta$ and EW(H$\beta$) falls within the wavelength range of the JWST/NIRSpec spectrograph for $z\approx 4$--9. The use of higher-order Balmer lines instead of H$\beta$ (H$\gamma$, H$\delta$ etc.) would extend the range to $z\gtrsim 9$, but at the expense of losing the [OIII]$\lambda$5007 line, which is useful for getting a handle on the metallicity (Sect.~\ref{metallicity}). Since the proposed method is applicable for relatively blue SEDs ($\beta\lesssim -2.3$) and positive EW(H$\beta$), the most challenging aspect of these observations is to measure the $H\beta$ continuum. Using the JWST/NIRSpec exposure time calculator version P1.5 with the $R=100$ setting, we estimate that one should be able to get a $S/N\sim 4$ measurement of the H$\beta$ continuum (giving an error $\sigma(\log_{10}\mathrm{EW(H\beta)})\approx 0.1$, since the error on the line flux will be negligible by comparison) by integrating over $\lesssim 50$ \AA{} bins in the rest-frame spectrum) in 10 h for a 100 Myr old, $Z=0.020$, $f_\mathrm{esc}=0.5$, constant-SFR galaxy with stellar mass $3\times 10^{8}\ M_\odot$ at $z=7$. At $z=9$, the corresponding mass limit would be $1\times 10^{9}\ M_\odot$.

Galaxies this massive are well within the detection limits of ultradeep Hubble Space Telescope imaging, and many $\sim 10^8$--$10^9\ M_\odot$ objects have already been discovered in current $z\gtrsim 7$ samples \citep[e.g.][]{Schaerer & de Barros b, McLure et al. b}. Strong lensing by foreground galaxy clusters may boost the fluxes of high-redshift background objects by magnification factors of up to $\mu\approx 30$--100 \citep[e.g.][]{Johansson et al.,Zackrisson et al. c,Coe et al.}. By targeting lensed fields, it should therefore be possible to push these detection limits down to $\approx 1\times10^7\ M_\odot$ ($z=7$) and $\approx 3\times 10^7$ ($z=9$), if $\mu\approx 30$ is assumed. In terms of the rest-frame 1500 \AA{} luminosities often used when measuring the luminosity function at high redshifts, this corresponds to $M_{1500}\lesssim -16.0$ at $z=7$ and $M_{1500}\lesssim -17.5$ at $z=9$. These detection limits convert into star-formation rates of $\gtrsim 0.1\ M_\odot \ \mathrm{yr}^{-1}$, placing them slightly above the level where stochastic sampling of the stellar initial mass function starts to have serious effects on the LyC flux of galaxies \citep{Forero-Romero & Dijkstra}.  

On a somewhat longer timescale, similar observations will also be possible with groundbased telescopes like the {\it Giant Magellan Telescope}\footnote{http://www.gmto.org/}, the {\it Thirty Meter Telescope}\footnote{http://www.tmt.org/} and the {\it European Extremely Large Telescope}\footnote{http://www.eso.org/sci/facilities/eelt/}.

As discussed in Sect.~\ref{attenuation}, complementary observations of the rest-frame mid/far-IR with SPICA or ALMA could in principle provide information on the fraction of LyC photons destroyed directly by dust. Such observations could also reveal the presence of heavily embedded components within the target galaxies. Formally, the $f_\mathrm{esc}$ derived from JWST observations can only reflect the star formation that actually contributes to the rest-frame UV/optical SED, and extremely dust-obscured star formation (undetectable in the optical, but bright at IR wavelengths, as in infrared-luminous galaxies at lower redshifts; e.g. \citealt{Choi et al.}) would shift this $f_\mathrm{esc}$ estimate away from that relevant for simulations of galaxy-driven reionization, where the {\it total} star formation rate of galaxies needs to be converted into a LyC flux emitted into the IGM. Based on the dust SEDs for star-forming dwarf galaxies presented by \citet{Takeuchi et al.}, we estimate that strongly lensed ($\mu\approx 100$), $z\approx 6$--9 galaxies with SFR $\sim 10\ M_\odot$ yr$^{-1}$ should be detectable with ALMA in $\sim 10$ h exposures at $\approx 400$--$1000\ \mu$m. SPICA may be able to detect such lensed $z\approx 6$ galaxies at $\approx 80\ \mu$m in less than 1 h, but this requires that one can push the photometry a factor $\approx 5$ below the formal SPICA confusion limit -- for instance by using auxiliary, high-resolution data from other wavelength bands (e.g. obtained with JWST or ALMA) to subtract off the flux contribution from nearby interlopers.

\subsection{Spectroscopic versus photometric signatures of LyC leakage}
In its current form, the proposed method can be used to assess $f_\mathrm{esc}$ for the brightest galaxies at $z\gtrsim 6$, but is likely to provide interesting limits for galaxies exhibting very high LyC escape fractions only ($f_\mathrm{esc}\gtrsim 0.5$). However, the rest-frame UV/optical SED carries more information than contained in the EW(H$\beta$)-$\beta$ diagnostic, and stronger constraints can possibly be set by making use of the entire SED obtained from JWST/NIRSpec observations. In future papers, we intend to explore the true $f_\mathrm{esc}$ limits that can be obtained from spectroscopy using more detailed SED simulations. While it is likely that $f_\mathrm{esc}$ estimates for less extreme leakers may be obtained this way, the luminosity limits of the method are unlikely to change significantly. Hence, only objects with $M_{1500}\lesssim -16.0$ (stellar masses $\gtrsim 10^7\ M_\odot$) can be probed this way. 

Galaxies in this mass range may be insufficient to explain the ionization state of the Universe at $z\lesssim 9$ \citep[e.g.][]{Finkelstein et al. b,Ferrara & Loeb,Alvarez et al.,Paardekooper et al.}, and methods based on photometry rather than spectroscopy (in the vein of \citealt{Ono et al.}, \citealt{Bergvall et al. b}, \citealt{Pirzkal et al. a,Pirzkal et al. b}) will be necessary to estimate $f_\mathrm{esc}$ for objects at lower masses. While photometric methods are unlikely to provide strong constraints for individual targets, they can on the other hand be applied to larger samples of galaxies. In the future, we therefore also intend to investigate the likely $f_\mathrm{esc}$ limits that can be set using photometry with JWST/NIRCam and JWST/MIRI filters.

\subsection{Population III galaxies}
\label{popIII}
\begin{figure}
\plotone{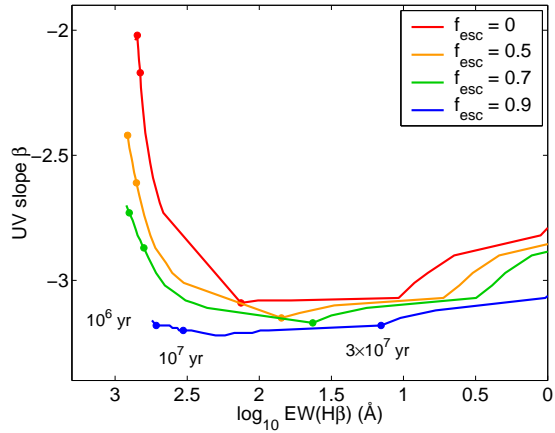}
\caption{EW(H$\beta$) versus $\beta$ for population III ($Z=0$) galaxies with a top-heavy IMF (characteristic stellar mass $\sim 10\ M_\odot$). A short burst of constant SFR (lasting $10^7$ yr) is assumed, after which the SFR drops to zero. Ages of 1, 10 and 30 Myr are indicated by markers along each track. The different line colours represent different LyC escape fractions: $f_\mathrm{esc}=0$ (red),  $f_\mathrm{esc}=0.5$ (orange),  $f_\mathrm{esc}=0.7$ (green) and  $f_\mathrm{esc}=0.9$ (blue). Due to the strong impact of nebular emission on the UV slope $\beta$, the evolutionary tracks look very different from those of high-metallicity galaxies, but since the tracks remain separated for the first $\sim 10$ Myr (roughly the time during which population III galaxies can be identified based on JWST colours; \citealt{Zackrisson et al. b}), it should be possible to assess $f_\mathrm{esc}$ from a simulataneous measurement of EW(H$\beta$) and $\beta$ even for these exotic objects.  
\label{PopIII_fig}}
\end{figure}

Population III galaxies may exist at high redshifts \citep[e.g.][]{Stiavelli & Trenti} and are predicted to have high $f_\mathrm{esc}$ \citep{Johnson et al.,Benson et al.}. Such objects may be detectable with JWST in strongly lensed fields \citep{Zackrisson et al. c} and should be identifiable because of their unusual colours \citep{Inoue d,Zackrisson et al. b,Zackrisson et al. d} and spectra \citep[e.g.][]{Schaerer,Inoue c,Inoue d,Zackrisson et al. b}, at least for ages up to $\sim 10^7$ yr \citep{Zackrisson et al. b}. While the low masses predicted for such galaxies (total stellar mass perhaps no more than $\sim 10^4\ M_\odot$; \citealt{Safranek-Shrader et al.}) makes it unlikely that these objects would contribute substantially to the reionization of the Universe, the spectral diagnostics we propose could in principle be used to assess $f_\mathrm{esc}$ for these galaxies as well. This is demonstrated in Fig.~\ref{PopIII_fig}, where we show the evolution predicted in the EW(H$\beta$)-$\beta$ diagram for a population III galaxy experiencing a brief burst ($10^7$ yr) of zero-metallicity star formation, in the case of a radiation-bounded nebula with holes. The characteristic masses of population III star formation is generically predicted to be top-heavy, but the exact stellar IMF remains highly uncertain. Here, we have used the \citet{Raiter et al.} stellar SEDs for a log-normal IMF with characteristic mass $10\ M_\odot$, dispersion $1\ M_\odot$ and wings extending from 1--500 $M_\odot$, but the results are similar for other IMFs explored by \citet{Zackrisson et al. b} during the early phase $\sim 10^7$ yr when population III galaxies should be uniquely identifiable. At low ages, the strong impact of nebular emission on the UV slope at these low metallicities makes the model tracks curve upward in the EW(H$\beta$)-$\beta$, but the tracks corresponding to different LyC escape fractions still remain clearly separated, thereby allowing an estimate of $f_\mathrm{esc}$. At higher ages ($\gtrsim 10^7$ yr), EW(H$\beta$) drops while $\beta$ evolves very little, which causes substantial degeneracies between $f_\mathrm{esc}$ and the IMF (not shown, to avoid cluttering).   

\subsection{Impact on 21-cm studies of reionization}
\label{reionization}
One of the most important problems in observational cosmology is identifying the sources that reionized the Universe. It is generally assumed that a substantial fraction of the ionizing photons were produced by galaxies, but other objects such as quasars \citep[e.g.][]{dijkstra2004}, mini-quasars \citep[e.g.][]{madau2004}, micro-quasars \citep[e.g.][]{Mirabel} and population III stars \citep[e.g.][]{venkatesan2003} may also have played a role. Currently, there exist mostly indirect constraints on the timing of the reionization epoch. The spectra of high-redshifts quasars suggest that reionization was completed by $z\approx 6$ \citep{Fan et al., mortlock2011}, while observations of the cosmic microwave background can be used to put limits on the extent of the reionization period \citep[e.g.][]{Komatsu et al.,larson2011,Pandolfi et al.,Zahn et al.}. The detailed reionization history at $z\gtrsim 6.5$ remains highly uncertain, with statistics of Ly$\alpha$ emitters seemingly suggesting that the IGM was still highly neutral at $z\approx 6.5$--$7$ (e.g.\ \citealt{ouchi2010,pentericci2011,jensen2013}), while IGM temperature measurements instead indicate that reionization was already mostly complete by these redshifts \citep{theuns2002,raskutti2012}.

The most promising prospect for constraining the reionization history in the near future is by observing the 21-cm line emission from the partly neutral IGM. Radio observatories such as LOFAR (van Haarlem et al. 2013, subm.), MWA \citep{tingay2012} and PAPER \citep{parsons2010} are aiming for a detection at some point within the next few years. The 21-cm signal encodes a wealth of information about the physics of the reionization epoch, but the first generation of experiments will be focused on simple statistics such as the variance and power spectrum of the 21-cm brightness temperature. These statistics will not be able to constrain the evolution of the ionized fraction directly, but will have to be interpreted by comparing to simulations (e.g. \citealt{mcquinn2007,lidz2008,iliev2012}). 

Most simulations to date include only galaxies as sources of ionizing photons, and use simplistic recipes to assign ionizing fluxes to the sources, tuned to reproduce the available observational constraints. With measurements of the LyC escape fraction in combination with observed UV luminosity functions, much of the ``wiggle room'' for these recipes would be eliminated. This in turn would make 21-cm measurements more powerful for constraining the sources of reionization by enabling better comparison between the 21-cm statistics predicted by different source models.

\section{Summary}
\label{summary}
If galaxies were responsible for the reionziation of the Universe, significant LyC photon escape from these objects must have occurred at $z\gtrsim 6$. Our ability to put this scenario to the test is hampered by the neutral IGM at these epochs, which precludes a direct measurement of the leaking LyC flux. However, since the LyC escape fraction $f_\mathrm{esc}$ regulates the relative impact of nebular emission on the SEDs of galaxies, indirect information on this parameter should be retrievable from the SED. As a first demonstration of this, we argue that extreme LyC leakers (with $f_\mathrm{esc}\gtrsim 0.5$) may be identifiable from spectroscopic measurements of the equivalent width EW(H$\beta$) of the H$\beta$ emission line and the UV slope $\beta$. By targeting strongly lensed galaxies, JWST/NIRSpec spectroscopy should allow this technique to be applied to galaxies with stellar masses $\gtrsim 10^7\ M_\odot$ and redshifts up to $z\approx 9$. We explore the impact of metallicity, star formation history and dust extinction on the EW($H\beta$) and $\beta$ diagnostics, and while all of these effects may complicate $f_\mathrm{esc}$ estimates to some extent, both the metallicity and the star formation history can be constrained by other spectral features that are readily detectable by JWST/NIRSpec. Dust represents a more troublesome issue, since the rest-frame UV/optical SED accessible to JWST/NIRSpec at $z\approx 6$--9 may contain insufficient information to properly gauge this effect. While several studies have indicated that Lyman-break galaxies in the reionization epoch have suffered very little dust reddening, direct absorption of LyC photons by dust may still occur, with lingering uncertainties in the inferred LyC escape fraction as a result. Heavily obscured star formation that contributes very little to the rest-frame UV/optical may in principle also bias $f_\mathrm{esc}$ estimates based on this part of the SED. We suggest that supplementary observations of lensed $z\approx 6$--9 galaxies with ALMA or the planned SPICA mission may help constrain such scenarios by measuring the dust emission peak in the rest-frame mid/far-IR.

\acknowledgments
E.Z acknowledges funding from the Swedish National Space Board and the Swedish Research Council. A.K.I. acknowledges funding from the Ministry of Education, Culture, Sports, Science, and Technology (MEXT) of Japan (KAKENHI: 23684010). The anonymous referee is acknowledged for helpful comments on the manuscript.\vspace{5mm}

\end{document}